# Energy conservation in dissipative processes:
# Teacher expectations and strategies associated with imperceptible thermal energy

Abigail R. Daane,[1] Sarah B. McKagan,[1,2] Stamatis Vokos,[1] and Rachel E. Scherr[1]
1. Department of Physics, Seattle Pacific University, Seattle, WA 98119
2. American Association of Physics Teachers, College Park, MD 20740

*Abstract:* Research has demonstrated that many students and some teachers do not consistently apply the conservation of energy principle when analyzing mechanical scenarios. In observing elementary and secondary teachers engaged in learning activities that require tracking and conserving energy, we find that challenges to energy conservation often arise in dissipative scenarios in which kinetic energy transforms into thermal energy (e.g., a ball rolls to a stop). We find that teachers expect that when they can see the motion associated with kinetic energy, they should be able to perceive the warmth associated with thermal energy. Their expectations are violated when the warmth produced is imperceptible. In these cases, teachers reject the idea that the kinetic energy transforms to thermal energy. Our observations suggest that apparent difficulties with energy conservation may have their roots in a strong and productive association between forms of energy and their perceptible indicators. We see teachers resolve these challenges by relating the original scenario to an exaggerated version in which the dissipated thermal energy is associated with perceptible warmth. Using these exaggerations, teachers infer that thermal energy is present to a lesser degree in the original scenario. They use this exaggeration strategy to productively track and conserve energy in dissipative scenarios.

## I. INTRODUCTION

The Next Generation Science Standards [1] emphasize the importance of tracking and conserving energy through physical scenarios.[1] A critical component of tracking and conserving energy is the recognition of the forms of energy present during a scenario. Forms of energy are generally identified by a *perceptible indicator*, such as motion, sound, height, or warmth, that provides sensory evidence for the presence of energy. In a rollercoaster scenario, for example, changes in height and speed of the rollercoaster are the perceptible indicators used to track energy as it transforms from gravitational energy to kinetic energy.

This method of tracking energy by its perceptible indicators is particularly useful in idealized scenarios that neglect dissipative processes (e.g., a rollercoaster moving on a frictionless track). These are the kinds of scenarios most often emphasized in physics courses. In the case of a real rollercoaster, gravitational energy does not all end up as kinetic energy; some ends up as thermal energy in the rollercoaster, the track, and the surrounding air. We observe that learners who engage with such dissipative processes recognize changes in energy associated with perceptible indicators (e.g., changes in gravitational energy associated with changes in the height of a rollercoaster), but often do not identify changes in energy associated with imperceptible indicators (e.g., the production of thermal energy in a scenario in which the rollercoaster doesn't feel hotter). The disappearance of perceptible indicators can seem to contradict the energy conservation principle. This strong association between forms of energy and perceptible indicators may account for some of the student difficulties described in previous research on applying energy conservation to everyday phenomena (e.g., [2]). Further, we find that this association leads to concern and puzzlement even for learners who do not have "difficulties" with energy conservation in the traditional sense.

---

[1] The term *scenario* refers to an "energy story" involving the objects comprising the system that has a predetermined time development (e.g., a basketball rolls to a stop, or an incandescent bulb glows steadily) [4].

Our observations of learners discussing dissipative scenarios in K-12 teacher professional development and high school classrooms have led us to better understand expectations learners have about energy transfers and transformations. We have also identified productive strategies that teachers-as-learners employ in successfully tracking and conserving energy through dissipative processes. In this paper, we make the following claims about learners' ideas regarding energy conservation in dissipative processes:

1. Learners expect that energy associated with a perceptible indicator will also be associated with another perceptible indicator when the energy transforms. In particular, learners expect that kinetic energy associated with *visible* motion will transform into thermal energy associated with *palpable* warmth. This expectation challenges their commitment to energy conservation when all energy indicators disappear from perception.
2. Learners accept the presence of thermal energy associated with the imperceptible indicator of warmth when they recognize that warmth would be perceptible in an exaggerated scenario. For example, learners accept the presence of thermal energy in a rollercoaster scenario when they recognize that warmth is perceptible in a space-shuttle re-entry scenario.

We support these claims by first describing the physics of energy dissipation and the perceptibility of indicators of energy forms (Section II). We then review previous research on learning about energy conservation (Section III) and introduce the context in which our research takes place (Section IV). Next, we share evidence of learners' expectations about perceptible indicators of energy as well as strategies that support their acceptance of imperceptible thermal energy (Sections V and VI, respectively). The significance of these results and the instructional implications are described in Section VII.

## II. PHYSICS OF ENERGY DISSIPATION

Energy dissipation, as discussed in this paper, is the process of macroscopic kinetic energy transforming into thermal (or internal) energy through interactions among microscopic particles that randomize their motion and position and spread energy more uniformly throughout a system. Dissipated energy is sometimes described as "energy lost from an open system" [3], where "lost" energy indicates energy that is degraded, or cannot be used for the performance of work [3, 4]. The NGSS, to which teachers are accountable, does not explicitly require understanding of energy dissipation [1]. However, the NGSS's primary learning goals about energy – that it is conserved, that it manifests in multiple ways, and that is continually transferred from one object to another and transformed among its various forms – require accounting for energy wherever it goes in the scenario of interest. Further, the NGSS's emphasis on energy-efficient solutions to societal problems is reflected in its statements about scenarios involving "diffuse energy in the environment," usually in the form of thermal energy. Though the NGSS refers more to processes of conduction than dissipation (e.g., "When machines or animals 'use' energy, most often the energy is transferred to heat the surrounding environment"), dissipation is a significant feature of energy scenarios that embody NGSS priorities.

In many energy scenarios occurring near room temperature, the thermal energy produced by dissipation cannot be perceived by human senses (we cannot feel any indication of the energy's presence). For example, when a ball rolls to a stop, the motion associated with the ball's kinetic energy disappears and the warmth associated with the thermal energy produced in the ball, air, and ground is likely to be imperceptible. As humans, the disappearance of perceptible indicators for energy leads to a contradiction between what we experience and what we expect to experience. Our intuition supports the assumption that sensory experiences have certain common dimensions that transcend specific modalities of the senses: "for example, bright is like loud because both are intense… In this view, then, the reason that brighter lights are perceived to be like louder sounds is because they share a common property, intensity… Bright and loud are conceptually understood as being about some amount of physical energy" [5]. It follows that a person who accepts that energy is conserved would also expect that the perceptibility of that energy's indicators to be "conserved." For example, in the scenario of a ball rolling to a stop, the disappearance of a perceptible indicator (motion of the ball) without replacement by another perceptible indicator can seem to suggest the disappearance of energy and a violation of the principle of energy conservation.

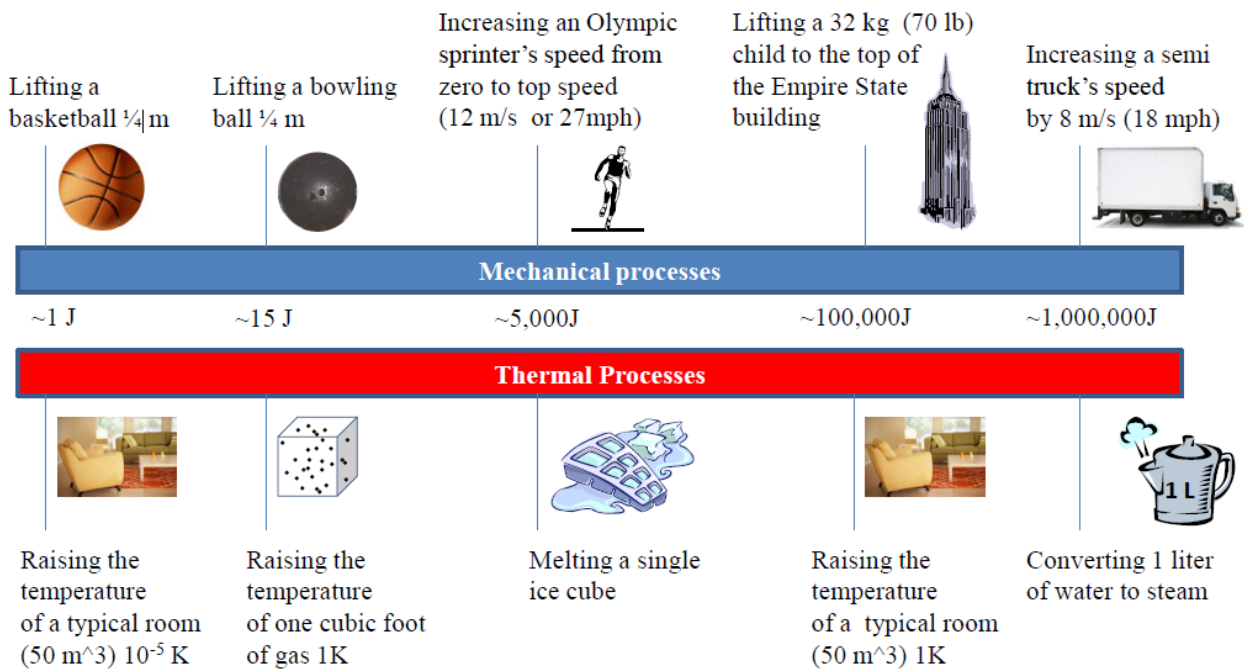

**Figure 1:** A comparison of the energy associated with various thermal and mechanical processes.

In many cases, energy associated with a perceptible indicator will *not* be associated with another perceptible indicator when it transforms to another form. Figure 1 shows examples of thermal and mechanical processes requiring varying amounts of energy. Changes in mechanical energy of about one joule may be associated with easily perceptible indicators (e.g., lifting a basketball ¼ m), but if all of that energy were transformed to thermal energy, it would only increase the temperature of a typical room (50 cubic meters) by an imperceptible $10^{-5}$ K ($10^{-5}$ °F). To produce an easily perceptible quantity of thermal energy, such as that associated with raising the temperature of a typical room from 40 to 60°F, one would need to drop almost 190,000 basketballs from a height of 1 meter. The difference in the perceptibility of energy indicators for various forms can cause learners to struggle with tracking energy in dissipative processes.

### III. PRIOR RESEARCH ON LEARNING ABOUT ENERGY IN DISSIPATIVE PROCESSES

The majority of research analyzing student understanding of energy in dissipative processes has appeared, almost entirely implicitly, in research focused on student understanding of the conservation of energy principle. Many of these studies use physics scenarios that involve dissipative processes (or idealized physics scenarios that would involve dissipation in the real world). For example, one study uses a car that coasts to a stop and a golf ball that is hit and bounces several times, reaching a smaller and smaller height before coming to a stop [6]. Other studies use a damped swinging pendulum [7, 8]. Another uses a ball rolling up and down the sides of a bowl, asking students to neglect frictional effects [9]. These scenarios, in the real world, all involve a decrease in total kinetic and potential energy and a compensating yet imperceptible increase in thermal energy (e.g., as a pendulum slows to a stop, it does not feel warmer).

The general consensus of this research, across a variety of contexts, is that many students and some teachers have difficulty understanding and applying energy conservation [2, 6-16]. One study explicitly describes the transformation from kinetic to thermal energy as problematic in secondary education: interviews with 34 German students (15-16 years old) reveal that after physics instruction, students "have difficulties in using the idea of the transformation of kinetic energy to heat energy to explain relevant processes" [7, p. 99]. In a scenario in which a pendulum swings to a stop, only four out of 34 students described kinetic energy as transforming into thermal energy; the rest of the student responses were attributed to a lack of understanding of energy conservation.

Another way in which students and some teachers appear to contradict the conservation of energy principle is to describe the energy in dissipative thermal processes as being used up or lost [6-8, 12, 15]. For example, one British student explained her thinking about the energy conservation principle as it applies to the process of a lamp shining in this way:

> "That principle of conservation, Miss, I don't believe it. You know when you have a battery and a lamp, and the battery has electrical energy, right? And it goes to heat and light in the lamp. Well, I mean, the heat evaporates and the light goes dim. So the energy has gone. It isn't there is it?" [6]

A similar finding appeared in a study in which many university introductory biology students were "unable to apply the idea of energy conservation" to biological settings even though almost 98% of them identified the correct statement of the conservation of energy principle [12]. Some "used the terms used up, created, made or lost in their explanations [of energy processes]" [12]. When students were asked to identify incorrect phrases in a number of sentences describing dissipative processes, "only 4% of the students in the whole group correctly underlined *used up* as an incorrect phrase and wrote in the scientifically acceptable phrase, *converted to different forms*" [12]. In our own earlier work, we argued that the idea that energy is used up or lost can be productively aligned with the concept of energy degradation [4]. In this paper, we focus on the challenge to energy conservation that is presented when thermal energy indicators are imperceptible.

Other research found that when students conserve energy in dissipative processes, they sometimes mistakenly describe kinetic energy as transforming into potential energy instead of thermal energy [6, 8]. For example, British high school students analyzed the energy at the end of a scenario in which a golf ball bounces to a stop. Rather than describing the energy as dissipated, students claimed that the stopped ball had "stored up" the energy, and that the energy could be used again [6]. University students in the U.S. came to a similar conclusion when asked about a damped pendulum: they described the kinetic energy of the pendulum as transforming into potential energy as the pendulum slowed to a stop [8]. Their response shares features with a canonical account of the energy dynamics of the scenario: it respects the principle of energy conservation by inferring a transformation into a form of energy with no perceptible indicator. However, their response misconstrues "potential energy" as entirely hidden or latent [17], rather than associated with the configuration of interacting objects.

All of these studies characterize students as having difficulty understanding and applying energy conservation without mentioning the possible role of imperceptible energy indicators in dissipative processes. We take as a premise that learners at all levels have rich stores of intuitions about the physical world, informed by personal experience, cultural participation, schooling, and other knowledge-building activities [18-20]. Some of these intuitions are "productive," meaning that they align at least in part with disciplinary norms in the sciences, as judged by disciplinary experts [21, 22]. Learners may only apply these intuitions episodically: at some moments of conversation with instructors and peers there may be evidence of productive ideas, whereas at other moments productive ideas may not be visible [23]. This perspective suggests that rather than having a "difficulty" or a "misconception" about conservation of energy, the learners in our study are attempting to reconcile understanding of the conservation of energy principle with their intuition that energy indicators should remain perceptible as energy transforms. Our work here aims to build on and reframe previous research about difficulties with energy conservation, showing that learners' intuitions about perceptibility can be used productively to support a greater understanding of energy conservation.

## IV. RESEARCH CONTEXT

### A. Research Methods

This paper reports on a phenomenological study using data gathered by the Energy Project, a six-year NSF grant focused on the teaching and learning of energy. As part of the Energy Project, a variety of classrooms were observed in an effort to better understand how learners view and apply energy concepts. "Learners" is a broad term that we use to refer to three populations: (1) elementary and (2) secondary teachers-as-learners in summer professional development courses held at Seattle Pacific University, and (3) students in high school science courses taught by some of these teachers. Observations of learners' discussions in these three contexts promoted the investigation of the following two research questions:

1. What challenges learners' commitment to energy conservation in dissipative processes?
2. What instructional strategies can help address the challenge that energy dissipation presents to the law of energy conservation?

We found examples of this challenge across these diverse groups, suggesting that certain intuitions and understandings of dissipative processes are common to a variety of different learners.

Researchers collaborating with the Physics Education Research Group at Seattle Pacific University observed professional development courses and recorded their observations in real time using field notes, photography, artifact collection (including written assessments and teacher reflections) and video recordings for each observation. In these courses, teachers generally worked in groups of 3-4, with 4-8 groups in each class; two groups were recorded daily. In real time, researchers identified particular moments of interest and marked them for later analysis. Later, researchers chose episodes[2] that addressed the phenomenon of interest. For this analysis, video episodes were identified through (1) initial observations by videographers and (2) a search for key terms in the field notes which could relate to energy dissipation (e.g., dissipation, disappear, missing, spreading, diffusion, thermal energy). Episodes were selected when learners made visible the challenge to energy conservation. In each selected episode, learners articulated in some way the lack of evidence of the presence of energy, often asking "where did the energy go?" or describing the energy as disappearing. The groups in these episodes worked to solve this challenge for the remainder of the discussions. Detailed transcripts and narratives of each episode were produced and corroborated by multiple viewings from multiple researchers. A group of researchers then collaboratively analyzed several aspects of communication including gestures, facial expressions, interactions between participants, bodily behavior, and the context in which the activities occur [24, 25]. Fifteen episodes from six distinct discussions were isolated and captioned to illustrate learner engagement with issues of imperceptibility of thermal energy in dissipative processes. These episodes are described in Sections V and VI.

### B. Instructional Context

Instructors of both the professional development courses and the high school science courses in this study use Energy Tracking Representations to support learners in thinking about energy scenarios. These representations promote energy conservation and tracking in real-world scenarios [4, 26-31]. One of the representations used in all courses is an embodied learning activity called Energy Theater [31]. The rules of Energy Theater are:

---

[2] We use the term "episode" to refer to a video-recorded stretch of interaction that coheres in some manner that is meaningful to the participants [24].

- Each person is a unit of energy in the scenario.
- Regions on the floor correspond to objects in the scenario.
- Each person has one form of energy at a time.
- Each person indicates their form of energy in some way, often with a hand sign.
- People move from one region to another as energy is transferred, and change hand sign as energy changes form.
- The number of people in a region or making a particular hand sign corresponds to the quantity of energy in a certain object or of a particular form, respectively.

An Energy Theater enactment illustrates a group's shared understanding of the energy scenario. For example, a group of teachers-as-learners shown in Fig. 2 analyzes the scenario of a ball being lowered at constant velocity by a person. This group's Energy Theater enactment begins with the configuration shown in the figure: Four teachers represent gravitational energy in the ball, standing in a region on the floor representing the ball with their hands raised over their heads. Two teachers represent chemical energy in the person, using a sandwich-eating gesture (making a chewing motion with their hands holding an imaginary sandwich near the mouth). Finally, two teachers represent kinetic energy, one located in the ball and one located in the person, by their own fists circling each other in front of their stomachs. As they act out the scenario, the teachers representing gravitational energy in the ball and chemical energy in the person each transform into kinetic energy and then into thermal energy. The two teachers representing kinetic energy do not change form or move to another location.

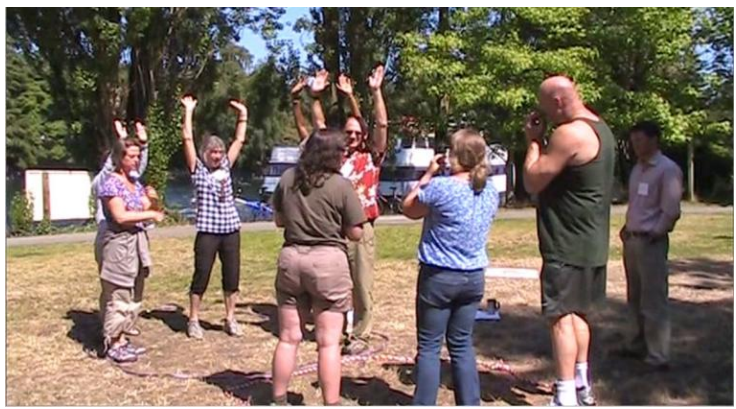

**Figure 2:** Teachers in a secondary science professional development course perform Energy Theater for a ball being lowered at constant velocity.

## V. REJECTIONS OF THERMAL ENERGY IN DISSIPATIVE PROCESSES

In this section, we present data supporting our assertion that learners expect that energy associated with a perceptible indicator will be associated with another perceptible indicator when it transforms. In particular, we show that learners expect that kinetic energy associated with visible motion will transform into thermal energy associated with palpable warmth. We demonstrate this expectation by showing that learners initially reject ideas that violate it. Specifically, we show that learners reject suggestions that thermal energy is produced in dissipative processes. These rejections have been observed in all Energy Project professional development course levels and in high school classrooms, across a variety of dissipative scenarios (e.g., a ball being lowered at a constant velocity, water waves forming from the wind, an apple falling to the ground, a basketball rolling to a stop).

We categorize learners' rejections into four types, associated with varying degrees of adamancy. First, some teachers implicitly reject suggestions of thermal energy by ignoring thermal energy suggestions and continuing to search for perceptible energy indicators (Section V.A). Second, some teachers and high school students explicitly reject thermal energy as a possible product of a particular process (Section V. B).

Third, some teachers accept the idea that *some* thermal energy is produced, but reject the idea that *all* energy ends as thermal energy (Section V.C). Lastly, some teachers accept the production of thermal energy, but do so with skepticism and reluctance (Section V.D.). Each rejection of thermal energy suggests a violation of learners' expectation that perceptible indicators of energy should remain perceptible, thereby providing evidence for their commitment to energy conservation (Claim 1).

### A. Implicit Rejection

One common reaction to a suggestion of thermal energy in our professional development courses comes in the form of an *implicit* rejection, in which listeners do not respond to suggestions about thermal energy. Sometimes they discuss another topic, suggesting that they may not have heard or attended to the suggestion. Other times, they respond to a non-thermal aspect of the suggestion, showing that they heard the statement but are not prioritizing its thermal energy content. Below, we share three episodes from a 20-minute discussion about a ball lowered by a person at a constant velocity (the "lowering scenario"). In conversations like this one, which center on tracking and conserving energy, learners must first identify the initial form and amount of energy by some indicator (e.g., motion) and then track that energy through a process. This group of eight secondary teachers quickly notice and articulate discrepancies in energy indicators and then repeatedly ignore (or do not perceive, or decline to take up) suggestions of thermal energy. Their struggle to track the energy exemplifies the conflict between the teachers' commitment to energy conservation and their expectation that the indicators of energy should remain perceptible through a process.

In the first five minutes of this discussion, the teachers contrast the lowering scenario to other scenarios. Kate[3] focuses on the differences between the current scenario and dropping a ball to the ground. The comparison leads the group to articulate that (1) the gravitational energy of the ball and the person decreases, and (2) the kinetic energy of the ball does not increase. The first mention of thermal energy is in response to Kate, who asks where the people representing units of gravitational energy should go when they leave the ball:

| | |
|---|---|
| Jennifer: | Kinetic energy needs to be constant. |
| Kate: | So what happens to the gravitational energy? |
| Irene: | It's reduced. |
| Kate: | But where do you [gravitational energy units] go though? But where do you go? But where you gonna go? |
| Ted: | What are you going to transform to? Kinetic? |
| Irene: | We can't go to chemical energy. |
| Jennifer: | Why does it go to kinetic? It's not free falling. Somebody's holding it and moving it down. |
| Leah: | But it is losing GPE [gravitational energy] as you are closer to the ground. |
| Jennifer: | It's losing GPE, but does the GPE get converted toooo…? |
| Irene: | We need another word for the energy that's holding it. |
| Jennifer: | Could it be heat? |
| Kate: | Wait a sec! |
| Jennifer: | Because really |
| Kate: | [snap snap] Does the earth, is the earth now suddenly? |
| Jennifer: | No |
| Ted: | The energy, is it transferred to [the person]'s hands? Because he has to uh....He has to do work. He has to use more energy to slow the ball down. |
| Jennifer: | I think the GPE goes to heat. |
| Marta: | [to Ted] He has to do more to lower it than pushing it up? |

---

[3] All names are pseudonyms.

In this episode, Jennifer proposes that the energy is transformed into "heat" (or thermal energy[4]) twice in this first excerpt. After her first suggestion, Kate introduces the earth as a new object to consider. Ted and Marta focus instead on the role of the person's hand in lowering the ball. Although Jennifer's suggestion is clearly audible, no one in the group engages her idea, implicitly rejecting thermal energy. After 90 seconds of discussion of Marta's question (whether raising or lowering a bowling ball requires more work), Barry redirects the group to focus on the missing energy in this scenario.

Barry: We know that if the ball goes from here [raised position] to here [lowered position], GPE was reduced,
Ted: It had to go somewhere.
Barry: and where does that GPE go?
Jennifer: Thank you for boiling that down.
Ted: It's the same as the mousetrap [a previous scenario].
Irene: To the arms [gestures lifting hand weights]
Jennifer: You know what? Seriously, the only place it could be going is heat cause it's obviously not going anywhere useful. You know, the energy is not going back into [the person], energy can't go *into* the earth.
Irene: But what happens when you lift weights. It takes energy to hold the ball.
Jennifer: It's so much easier if we just have it be heat.
Barry: Hold on, Hold on, the GPE...
Debra: yah, I think I'm with you. [points to Jennifer]
Barry: Hold on, the GPE changes into KE [kinetic energy], but it just stays in the ball. It doesn't go anywhere after that.
Marta: Yah, but, we can't end up with more KE; it's constant.
Ted: So we're slowing down the GPE conversion into kinetic.

This time, instead of simply suggesting that the energy transforms into thermal energy, Jennifer argues for the production of thermal energy using a process of elimination. Thermal energy is the only option left: the energy "is not going anywhere useful," "is not going back into [the person]," and "can't go *into* the earth." Debra begins to voice support for Jennifer's idea after she states, "It's so much easier if we just have it be heat," but no one else acknowledges Jennifer's proposal.

None of the teachers in this exchange connect Irene's focus on lifting weights to Jennifer's argument for thermal energy. Instead, they consider Barry's suggestion that the energy stays in the ball as kinetic energy. Shortly thereafter, Leah directs the group back to the issue of the missing energy.

Leah: GPE needs to be getting fewer, and the kinetic needs to stay the same.
Jennifer: So then how about we have some of the people [units of energy] who are going from GPE to kinetic go away as heat or go into the earth or whatever you're....like they have to be, I mean
Irene: So in other words we need another circle [another rope to indicate the addition of the earth as an additional object]
Kate: Ok, consider you were the earth…

Once again, Jennifer offers a suggestion that thermal energy is produced, but this time accompanied by a possible location for the energy to end up (that the energy goes into the earth, an idea she dismissed in the last excerpt). Irene and Kate take up Jennifer's idea about the earth being involved, but do not address any content related to thermal energy. The group discusses the forces between the earth and the ball for about two minutes following this exchange.

---

[4] Learners (including secondary teachers) often use "heat" or "heat energy" to refer to a form of energy indicated by temperature (what we call thermal energy), rather than a transfer of energy driven by temperature difference (what we call heat) [29, 33-35]. An association of heat with the temperature of an object is common in everyday speech, in non-physics textbooks, and in standards documents [35, 36]. However, such an association is not aligned with disciplinary norms in physics, in which the energy associated with temperature is often termed ''thermal energy" or "internal energy," and in which the term "heat" refers to energy transfer from a body at higher temperature to one at lower temperature. Differentiation of heat and temperature was not a learning goal in the specific instructional sequences represented here.

In the first 10 minutes of the discussion of this scenario, Jennifer suggests the idea that kinetic energy transforms into thermal energy a total of five times. Because her statements are clearly audible, and because in several cases members of her group take up other parts of her statements, it seems unlikely that the other participants do not hear her. Instead, it seems that they do not respond to thermal energy as a compelling solution for their missing-energy problem. Because their rejection is implicit, there is little opportunity to infer reasons for their inattention to the thermal energy idea in these episodes. However, in the sections that follow, we can begin to infer the reasons from more explicit rejections.

### B. Explicit Rejection

We have observed explicit rejections of thermal energy in several courses. The first episode below is from the same group of teachers as above, and chronologically follows the previous episode. The next episode comes from a high school biology class.

#### i. "I don't think we need any heat."

After a series of implicit rejections of thermal energy described above, another teacher, Marta continues the discussion about the lowering scenario, suggesting that thermal energy is produced.

| | |
|---|---|
| Marta: | It [the amount of kinetic energy] should be the same, but the amount of GPE is decreasing. Let's just lose one person [unit of energy] to heat or something. |
| Barry: | I don't think we need any heat. |
| Marta: | Alright, but GPE is decreasing. |
| Jennifer: | What's happening over here [in the person] is that more food molecules are being converted to kinetic and then we're just going to say, to hell with heat! |
| Others: | [echo this sentiment] |
| Irene: | So do we need another circle [rope that represents an object] for the Earth? |

In this exchange, Marta repeats the observation that the gravitational potential energy (GPE) in the ball decreases and suggests that some of the energy is lost to heat. This time, Barry explicitly rejects the use of any thermal energy in the representation. Jennifer also explicitly rejects thermal energy as a solution, discarding her original ideas. Her new suggestion (to produce more kinetic energy in the person) is supported by the group, but this does not solve the problem of the missing energy.

For a few minutes the conversation continues to focus on where the energy has gone. Several teachers (first Ted, then Irene, Ted again, Leah, and Ted for a third time) repeat the observation that the energy indicators decrease and revoice the question, "Where did the energy go?" In so doing, they collectively maintain a firm commitment to both conserving and tracking the energy. However, they persist in their attempt to make the representation work without the imperceptible thermal energy.

#### ii. "The apple's not giving off heat."

Another example of an explicit rejection of thermal energy comes from a high school Advanced Placement/International Baccalaureate biology course[5] taught by a teacher who participated in our professional development. In this episode, eight 16-18 year old students participate in Energy Theater, discussing the scenario of an apple falling from a tree. Though the context is different, we observe the same struggle to identify thermal energy with imperceptible indicators while they work to conserve and track the energy in this activity. Prior to the following episode, the group assigns Lou, a senior student, to represent gravitational energy in the apple location as it hangs in the tree. They decide that Lou should transform (change hand signs) into kinetic energy as the apple falls. Another senior student, Aaron, asks the group what the kinetic energy in the apple (represented by Lou) should transform into as the apple hits the ground.

---
[5] The course is an advanced biology course that incorporates both Advanced Placement and International Baccalaureate curricula over a two-year period.

Aaron: Ok, the energy that Lou is right now [kinetic energy as the apple falls to the ground], he's being used by the apple, so he's not going to stay in there right?
Becky: He's not, the apple's not giving off heat.
Aaron: The apple, so then what happens with the kinetic energy? You can't stay in there.

Similar to teachers in our professional development courses, these high school students notice that some energy is unaccounted for and attempt to identify where it has gone. When Aaron recognizes that the kinetic energy in the apple is "being used" (which we interpret as "decreasing"), Becky responds with an unprompted and unexplained rejection of the production of thermal energy. Aaron asks the same question voiced in several of the above episodes: "What happens with the kinetic energy?" The students demonstrate their commitment to the principle of energy conservation in that they spend the majority of their remaining time striving to account for all of the energy.

In the end, these students do not identify thermal energy as the resulting energy form. Instead, they decide that Lou should act as potential energy after the apple hits the ground. Similar responses have been observed with university students discussing a swinging pendulum [8]. In that study, student responses were interpreted as indicating confusion between gravitational force and energy. Another possibility is that the students are using "potential energy" as a placeholder for an unidentified energy form, or any form of energy that is not associated with a perceptible indicator. Even without identifying the missing thermal energy, the students' use of potential energy shows a strong commitment to energy conservation within the scenario.

### C. Partial Rejection

A third form of rejection observed in our professional development is to reject the idea that all of the kinetic energy in a scenario could transform to thermal energy, but accept that some of it could transform (a partial rejection). We see many instances of this partial acceptance of thermal energy. For example, Marta (Section V.B.i) states that the group should "just lose one person [unit of energy] to heat," not accounting for all of the energy using thermal energy. In the examples below, teachers in both the secondary and elementary professional development courses reject the idea that all of the energy transforms into thermal energy.

Partial rejection of thermal energy production seems to align with the treatment of thermal energy in many traditional physics problems, in which some of the energy dissipates due to friction or drag. A possible counter-claim to our claim that learners reject thermal energy because of its imperceptibility is the claim that learners incorrectly believe that thermal energy is always small in amount, since physics examples often mention thermal energy in reference to friction and minimize or neglect it. However, as we will show in Section VI, teachers in our courses spontaneously bring up examples in which thermal energy dissipates in large quantities with perceptible indicators, suggesting that they do not believe that thermal energy from dissipation is always small.

i. **"I'm just saying all of it cannot be going into heat."**

Roland, a secondary teacher in the professional development course discussed above, participated in a different group's Energy Theater about the lowering scenario. That other group concluded that the energy all winds up as thermal, but Roland argues against that conclusion. He states, "I'm just saying all of it [the energy] cannot be going into heat." Roland suggests that instead, the energy might transform back into gravitational energy (similar to the conclusion made in Section V.B.ii). In this episode, he concedes that some thermal energy is produced, but continues to search for the remaining energy.

Roland: Okay, where does that energy go? Does all the kinetic - gravitational energy - potential, which has been turned into kinetic- You asked us, where does it go? We know that *some* of it goes into heat. Does it all go into heat or does some of it go somewhere else? That's the question to answer right?
[Digression in conversation about where the energy does not go]
Instructor: Where else, so Roland says, Roland says it could go to heat.
Roland: Well we know that some of it goes to heat.
Instructor: Well, does all of it-
Roland: I don't know that heat all of it- [shaking his head, no]
Instructor: Where else could it go besides heat?
Roland: I don't know ....where... Can I ask you that question?
Instructor: Sure! [drops a pen onto the table] Could you hear it?
Roland: Sound? Does, but is that much energy going into sound?
Instructor: I don't know.
[Digression about sound energy]
Roland: Well see? There, you have at least answered my question that all of it does not go into heat! [laughs] I did it! That's - I'm satisfied! I'm satisfied!

Roland repeatedly asks where the energy goes, agreeing that *some* of the energy transforms into thermal energy. However, he responds negatively to the idea that all of the energy "goes to heat." When the instructor suggests some of the energy transforms into sound energy, Roland expresses satisfaction that not all of the energy transformed into thermal energy. The fact that lowering scenario does not produce any audible sound is not discussed.

ii. **"It never made that much heat for all of us to be fanning!"**

In a professional development course designed for elementary teachers, we find participants engaging in similar struggles with imperceptible indicators of energy, even when the scenarios differ. In the following episode, a group of K-8 teachers focuses on a scenario in which a basketball rolls to a stop. At the beginning, Brice, a middle school teacher, convinces the group to review their current understanding of the energy scenario by enacting Energy Theater. He narrates as they act out the energy processes.

Brice: Some of the heat is in the floor from the friction. [Some teachers move from the location of the ball to the location of the floor, transforming from kinetic to thermal energy, indicated by fanning.] Okay.
Carrie: And then eventually you all have to stop moving [referring to the remaining teachers representing units of kinetic energy], so... do you turn into heat?
[Teachers in the group one by one transform into heat]
Mindy: Now you're the heat in the floor.
Brianna: It never made that much heat for all of us to be fanning!
Brice: Yeah but we're just little amounts of energy.
Bart: We're very small.
Carrie: Think the ball.
Brice: We're like atomically sized.
Carrie: You are the ball.
Brianna: Very small, very small.

In this enactment, thermal energy is represented by a fanning motion. When Brianna, an elementary teacher, sees the group enact all of the kinetic energy in the ball transforming into thermal energy, she exclaims, "It never made that much heat for us all to be fanning!" – i.e., she states that the scenario does not produce a large amount of thermal energy. Brice, Bart, and Carrie reassure her that the units of thermal energy are "very small."

The description of the energy units as being "very small" may contradict the rules of Energy Theater (and thus the principle of energy conservation) if the energy units are being described as smaller than they were before the energy transformed. In this interpretation, all four teachers may be seen as rejecting the idea that all energy has transformed into thermal energy, implicitly contradicting the principle of energy conservation. Alternatively, the teachers may be claiming that the total amount of energy in the scenario is very small (and conserved). Either way, the "small" size of the thermal energy units justifies the lack of perceptibility to them.

### D. Skeptical Acceptance without justification

In addition to the above types of rejection (ignoring, explicitly rejecting, or partially rejecting thermal energy), teachers sometimes accept the production of thermal energy skeptically. In some cases teachers state their inability to identify perceptible indicators or mechanisms for its production as a reason for their skepticism. At other times, they indicate that they are relying on thermal energy as a catch-all or last-resort explanation when no other account is forthcoming. In this section we return to the group of secondary teachers discussing the lowering scenario and the elementary teachers discussing the rolling-basketball scenario. We then share an episode from another elementary teacher professional development course.

#### i. Using thermal energy is "just like a Hail Mary pass"

After the secondary teachers discussing the lowering scenario from section V.B.i explicitly reject thermal energy, the group continues to talk through a series of questions about the missing energy and revisits the thermal energy suggestion.

Leah: I'm beginning to think that it [thermal energy] going to the air is a good idea. I really am-
Ted: That just seems so like,
Jennifer: like a giveaway.
Ted: It's just like a Hail Mary pass, it's just like I don't know, let's just go [throws an imaginary ball].

Leah's suggestion that thermal energy goes to the air is met with a rejection from Ted. He states that the production of thermal energy is "like a Hail Mary pass," using a term from American football for a long, low-probability throw made in desperation at the end of a game. In using this term and gesturing an aimless throw, Ted expresses a sense that this answer is a last-ditch attempt, unlikely to result in a successful outcome. Jennifer similarly describes Leah's suggestion as "a giveaway," as if thermal energy is the easy answer instead of the right one.

#### ii. Imperceptible energy indicators require "a leap of faith"

In the conversation among elementary teachers discussing the rolling-basketball scenario (Section V.C.i.) one teacher asks, "Why did it [the basketball] slow down? And where did the energy go, that would have kept it propelling at the same rate of speed?" The group reaches a consensus that some of the ball's kinetic energy is transformed into thermal energy. Brice accepts this conclusion, but also looks for other forms of energy, such as sound energy, to make up the rest.

Brice: So we've got this kind of energy [hand sign for kinetic] and we have this kind of energy [hand sign for thermal]. Is there any other kind? That's the question I'm asking. I don't think so, but...I mean is sound... is, you know, if you could measure the sound coming off the ball would that be a form of energy that's being lost, just like the heat energy?
[Brice's question is directed to the instructor, who redirects it to the group.]
Brianna: Heat led to stopping the ball. The sound didn't lead to stopping the ball.
Jack: Did we actually hear anything?
Adrienne: Not me. But then I wasn't paying attention.
Bart: But by that same token we can't have any of these other [inaudible word] because we've got no way of measuring the ba- the energy that was in the ball, we just assumed there was some. And then, the energy just went away.
Carrie: So it's all a leap of faith.

In response to the suggestion that sound energy is also produced, Bart argues that in tracking energy, there are limitations to what you can measure. He states that they "assumed" the energy was there and then it went away. Carrie responds to the group as a whole that "it's all a leap of faith," possibly in reference to the presence of energy at the end of this scenario where it seems to disappear. These teachers may be arguing that anything that is not measurable requires a leap of faith, rather than making an argument specific to thermal energy. In any case, Bart and Carrie state that they must rely on assumptions and a leap of faith to accept thermal energy as the solution in this scenario.

iii. **"I just have to say, okay, I believe it."**

In another elementary teacher professional development course, a small group of K-5 teachers discuss what happens to the energy of a vertically dropped object that hits the ground. The instructor describes bending a paperclip back and forth repeatedly and feeling the metal grow warmer. She uses this as an alternative, perceptible example of dissipation.

Instructor: There're some things, like when we did the paperclip, it seemed like we got a lot of heat out of very little motion.
Vicki: Heat out of a little bit of motion - That's interesting too!
Instructor: That's really interesting! You know, so.
Marissa: I think that is all the more reason, for me, the transfer of sound and heat is, I just have to say, okay, I believe it, because there is evidence in other ways like with the evidence, I can reason it, but with that I can't grasp what the evidence is. You know what I mean?

In her statement, Marissa explains that she must "just believe" that thermal and sound energy are present in certain scenarios because she "can't grasp what the evidence is" (i.e., she can't perceive warmth and sound). Earlier in the same conversation, Marissa expressed this concern by describing how she feels about her understanding of energy after it spreads into the atmosphere.

Marissa: I feel like once it gets to the air, atmosphere level, I have no conceptual understanding, and I know something happens, and that's where we got to the thermal discussion before-
Instructor: But you guys have been talking about that in terms of - so those guys have been moving against each other and bounding off one another in mass, what's happening at the molecular level?
Marissa: So we can *guess* that it's thermal.

Marissa expresses a concern that her lack of conceptual understanding leads her to "guess" that the result is thermal. Her doubtful acceptance is similar to Ted's "Hail Mary pass" and Carrie's "leap of faith" in the previous episodes. Marissa's statements are distinctive in that she claims that a lack of perceptible evidence limits her ability to reason about thermal energy.

E. Summary

The evidence above supports our claim that learners expect that energy associated with a perceptible indicator will be associated with another perceptible indicator when it transforms to another form. In particular, we have shown that learners expect that kinetic energy associated with visible motion will transform into thermal energy associated with palpable warmth. Evidence of this expectation is in the form of various degrees of rejection: Learners reject the idea that thermal energy is produced in scenarios in which warmth is not perceptible. We see these rejections from elementary teachers, secondary teachers, and secondary students in a variety of scenarios. These learners demonstrate a substantial commitment to the principle of energy conservation in that they strive to account for the kinetic energy that seems to have disappeared from the scenario. Our observations suggest that difficulties applying the conservation of energy principle to dissipative scenarios may have their roots in a strong association between forms of energy and their perceptible indicators.

We do not typically observe all four types of rejections and a successful identification of thermal energy in one conversation. However, the secondary teachers analyzing the lowering scenario articulate each of these reactions in a particularly illustrative conversation. Over the course of the 20 minute discussion, Jennifer and others suggest thermal energy as a possible solution seven times with various reasoning and all are rejected (see Figure 3). The reasoning used in the suggestion for thermal energy grows in substance as the conversation progresses: from no reasoning, to arguing that the energy is lost, to suggesting that thermal energy goes to the air, to recognizing the warming of the body of the person lowering the ball. As their energy reasoning becomes more sophisticated, the teachers engage more fully in explaining their reactions. They begin by rejecting thermal energy suggestions implicitly – ignoring the suggestion, changing the subject, or addressing a different idea unrelated to thermal energy. When Marta suggests heat (the next-to-last suggestion in Figure 3) the rejection becomes explicit. In her statement, she is partially rejecting thermal energy herself by only suggesting a small quantity of energy to transform. Finally, when Leah makes the suggestion shown last in Figure 3), Ted and Jennifer articulate that their skepticism stems from a lack of evidence for the transformation.

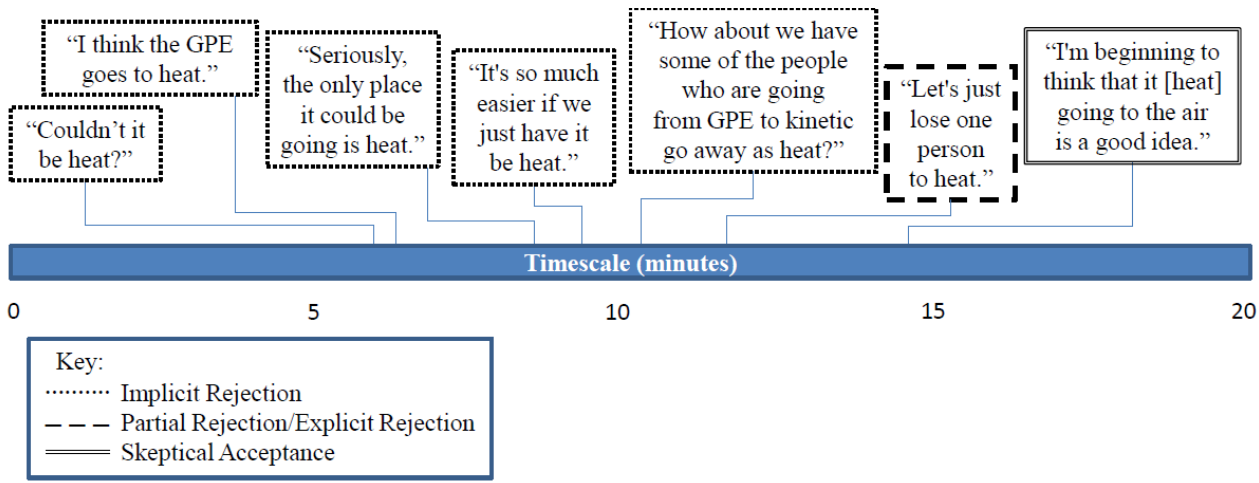

**Figure 3:** A timeline displays the suggestions for thermal energy as a resolution to the lowering scenario in chronological order.

We attribute the progress of this conversation partly to teachers' use of the Energy Theater representation. Energy Theater is designed to support teachers in conserving and tracking energy in complex physical processes [27, 31], including accounting for missing energy. Furthermore, Energy Theater's embodied action supports collaborative teams in theorizing mechanisms of energy transformation [29], including transformations from kinetic to thermal energy [32]. The development of the reasoning behind each suggestion of thermal energy and the teachers' investment in considering thermal energy highlights productive aspects of the Energy Theater activity and the resources that the teachers bring to the activity.

### VI. EXAGGERATION STRATEGY FOR JUSTIFYING THE PRESENCE OF THERMAL ENERGY IN DISSIPATIVE PROCESSES

Some teachers successfully resolve the issue of imperceptibility by using exaggeration (Claim 2). These teachers exaggerate the total amount of energy in a scenario so that the thermal energy becomes perceptible, then extrapolate back to infer the presence of thermal energy in the original scenario. Some teachers produce the exaggeration effect by imagining that the scenario repeats many times, building up the effects of the energy changes until those effects become perceptible (Section VI.A and VI.B). Other teachers relate the original dissipative scenario to an extreme version involving more total energy (Section VI.C). In all three episodes below, the exaggeration results in thermal energy that is either indicated by perceptible warmth or associated with burning.

### A. "Couldn't we just have his body heat up?"

The group of secondary teachers that analyzes the lowering scenario in the previous sections eventually uses an exaggeration strategy to identify the production of thermal energy. After Ted's "Hail Mary pass" statement (Section V.D.i), he begins to compare the energy in the lowering scenario to that in other related scenarios.

Ted: Hey wait a minute! Let's, let's reconsider something about, um, [the person]. So [the person] is holding the ball, he's holding the ball static and he's not moving. Okay so, the ball only has GPE right?
Leah: This is a good example.
Ted: So when so when he's lowering the ball at a constant speed, is there anything different about, is [the person] doing anything more, is he expending more energy than
Marta: That's what we were just, we said at the beginning.
Ted: He's just standing there and he's lowering the ball, I mean he's sort of collapsing himself down. [bends his knees and moves upper body back and forth] He's not really doing much and so it's, and all the ball is doing is-
Irene: But it's like weightlifting! [gestures lifting weights]
[Flare up of side conversations]
Jennifer: He's definitely expending more energy because he's engaging more of his body, like he went all the way down and I don't know if we're supposed to care or not but-
Ted: Uh, yah-
Jennifer: I would say from a chemical energy standpoint or whatever, that there's more involved.
Ted: But it's kinda like riding a bike down the hill, I mean, he's not really-
Irene: No it's not riding the bike because you are holding the ball!
Ted: But in terms just-
Irene: It's like weightlifting.
Jennifer: Hey wait a minute!
Ted: If he's just a robot or something-
Instructor: I'm not a robot.
Irene: But he's not!

Ted suggests that the lowering scenario might be comparable to a scenario involving a robot instead of a human, arguing that the person is "not really doing much." He also compares the scenario to riding (perhaps coasting) a bike down a hill. Irene asserts that the person's involvement is more comparable to weightlifting. Jennifer says that someone lowering a ball "is definitely expending more energy" than someone holding a ball in place, because he's "engaging more of his body." Jennifer's next suggestion may be related to the suggestion of metabolic activity:

Jennifer: Couldn't we just have his body heat up?
Irene: Exactly. *That's it!* [Shouts and points to Jennifer]
Debra: Thermal heat
Irene: *You've got it* because *so!* That's what I'm saying, you know it goes, the energy goes into the body! That's like weightlifting.
Ted: So the bowling ball has kinetic energy that gets transferred to [the person], and then he expends chemical, which then, turns into heat.
Irene: Makes you sweat.

Jennifer, Marta, and Leah had all suggested thermal energy in earlier parts of the conversation (see Figure 3), but their suggestions were not taken up. One possibility is that the group needed multiple opportunities to accept the idea. This interpretation is weakened by Irene's enthusiastic reception of Jennifer's current suggestion, as though it offered a *novel* solution to their shared puzzle. Another possibility is that the present suggestion is substantively different from the earlier ones. Jennifer had formerly suggested that the energy doesn't go anywhere useful, goes away as heat, or goes to the earth; Marta had proposed they lose one unit of energy to heat; and Leah had suggested the energy goes into the air. Jennifer's latest question, "Couldn't we just have his body heat up?" relates the transformation into thermal energy to a familiar physical experience, and suggests a metabolic mechanism. Irene elaborates the physical experience of metabolic effort with repetitive "weightlifting" gestures (bicep curls), suggesting that even if lowering a ball at a constant velocity *once* does not produce perceptible thermal energy, doing so repeatedly would "make you sweat." This repetition is the primary difference between the weightlifting scenario and the original lowering scenario. In other words, this group productively uses an exaggeration strategy to identify the production of thermal energy.

### B. "Same thing as doing a squat as slowly as you can"

In another professional development course for returning secondary teachers, Rita and Joe also use an exaggeration strategy to successfully locate missing thermal energy in the lowering scenario. Unlike the previous group of teachers, Rita and Joe first decide that the energy must transform into thermal energy, then work to justify the transformation.

Rita: If we are going to end up having T's [units of thermal energy] out here [points to the air/surroundings around the ball and person], then we need to account for it. 'Cause I think that is where some of it [the energy] goes. I mean, have you ever like, slowly lowered a ball, like a bowling ball in your hand, and you know, you're shaking? But it also takes "more energy" [gestures air quotes] to raise it that high.
Joe: Yah. Same thing as doing a squat as slowly as you can. It's hard.

Rita relates the original scenario to an exaggerated version in which the effort involved in lowering a ball causes "shaking." While she describes the shaking, she acts out the difficulty and effort it takes to lower an extra large, heavy bowling ball by shaking her hands and straining her voice as she lowers the imaginary ball. Joe refers to the experience of doing squats to emphasize that lowering the ball is "hard" to do. The bodily experiences of controlling motion and shaking are used as perceptible indicators of effort that justify the production of thermal energy. The exaggerations, expressed primarily in Rita's imitation of lowering an extremely heavy object, make it more plausible that the indicators of thermal energy can be perceptible and support the idea that lowering a ball with lesser effort also produces thermal energy.

### C. "We saw the space shuttle."

The same K-8 teachers who discuss the rolled basketball scenario (Sections V.C.ii. and V.D.ii) use exaggeration to justify the presence of thermal energy. Several of them agree that thermal energy is produced and seek to justify why they have settled on thermal energy.

Carrie: How do you know that it's being transferred into heat energy?
Bart: Because [the instructor] said so. Or someone like her.
Brice: Because we saw the space shuttle, coming through the atmosphere.

Bart offers skeptical acceptance without justification, similar to the teachers cited in Section V.D. Brice, however, compares the rolling-ball scenario to the extreme scenario of a space shuttle reentering the atmosphere (in which the production of thermal energy is dramatic and consequential). Here, the space shuttle is slowing to a stop through the atmosphere in a similar fashion to the basketball slowing to a stop on the ground.

## VII. CONCLUSION

The NGSS emphasize tracking and conserving energy through physical scenarios. In physics, we often track energy using perceptible indicators. However, in dissipative processes, the warmth associated with thermal energy is often imperceptible to human senses and its production goes unnoticed. We find that this imperceptibility of warmth counters learners' expectation that if energy is conserved, so too should the energy indicators be "conserved." The disappearance of perceptible indicators of energy can challenge learners' commitment to energy conservation by violating this expectation. We demonstrate this expectation by showing that learners engaged in tracking and conserving energy during Energy Theater initially reject ideas that violate this expectation. Learners react with some type of rejection of thermal energy, either implicitly, explicitly, partially, or by skeptical acceptance. We see these rejections from learners with different levels of background knowledge and in the context of a variety of scenarios. In many cases learners do not identify thermal energy as the final product in dissipative processes, aligning with the findings in previous literature. However, we believe that their intuition of associating perceptible indicators with particular forms of energy is productive. Teachers in our courses use an exaggeration strategy along with this intuition to imagine scenarios in which the perceptible warmth is created and successfully identify the production of thermal energy. We see this exaggeration strategy as a resource for supporting learners in better understanding the role of thermal energy in common scenarios and more readily accepting energy conservation. This resource was used by scientists in the demise of caloric theory, which emerged from Count Rumford's experiments with machine boring of cannon barrels: scientists recognized that the violent and seemingly inexhaustible increase in thermal energy in this exaggerated scenario could not have been resident previously within the cannon as caloric.

The issue of imperceptible energy indicators is not isolated to dissipative processes involving thermal energy. It can also arise in the production of sound energy, chemical energy, and other forms. We have seen teachers compare the same quantity of energy in two different forms and express surprise that the perceptible indicators and actual amount of energy are not necessarily correlated. For example, Vicki, an elementary teacher in a professional development course, stated, "I always think about all the sound in the city. I mean there's a tremendous amount! It seems intuitively like sound energy, what's it doing? …not much because nothing is heating up much! I mean, there's an apparent amount of a lot of energy sometimes that does very little in the end." Vicki describes a difference in what seems to her to be a large amount of sound energy and the relatively small amount of thermal energy for which she sees evidence in a city. Future work could investigate learners' expectations about perceptible indicators of a variety of forms.

In a world of increasing concerns about energy usage, vast amounts of dissipated thermal energy are produced in day-to-day activities. An emphasis on real world examples can give K-12 teachers and students the opportunity to think about issues of energy use, waste, and efficiency, highlighting the sociopolitical ramifications of the production of thermal energy. Ultimately, instructors can support learners in tracking and conserving energy by (1) using real-world examples that include dissipation, (2) encouraging learners to use the exaggeration strategy, and (3) explicitly contrasting the perceptibility of energy indicators across a variety of forms. Learners who resolve the mysterious loss of energy using exaggerated examples will be better equipped to understand energy conservation, more aware of their own limitations of perception, and more conscious of their own energy use in everyday situations.

## IX. ACKNOWLEDGEMENTS

We thank all the elementary and secondary teachers who have participated in Energy Project courses for their generosity in making their own and their students' reasoning accessible to the Energy Project team. We are grateful to Seattle Pacific University's Physics Education Research Group, including A. D. Robertson, L. S. DeWater, L. Seeley, and K. Gray for substantive discussions of this work. We also appreciate V. Sawtelle and P. Southey for their valuable feedback on the manuscript. This material is based upon work supported by the National Science Foundation under Grants No. 0822342 and 1222732.*Daane, McKagan, Vokos & Scherr*  *Energy conservation in dissipative processes*
*Page 17 of* 19